\documentclass[letterpaper,11pt]{article}

\usepackage{scrextend}
\usepackage{amsmath,amssymb,epsfig,bbm}
\usepackage[active]{srcltx}
\usepackage[all]{xypic}
\usepackage{MnSymbol}
\usepackage{slashed}
\usepackage{amsthm}

\usepackage{float}

\usepackage{color}

\usepackage{dsfont}

\definecolor{co}{cmyk}{0,0.7,0.3,0}
\definecolor{darkgreen}{cmyk}{1,0,1,.2}
\definecolor{m}{rgb}{1,0.1,1}

\newcommand{\be}{\begin{equation}}
\newcommand{\ba}{\begin{eqnarray}}
\newcommand{\ea}{\end{eqnarray}}
\newcommand{\nn}{\nonumber}

\def\d{\delta}

\def\oo{\omega}

\def\OO{\Omega}

\def\ca{{\cal A}}

\def\ch{{\cal H}}

\makeatletter
\newcommand{\eqnum}{\refstepcounter{equation}\textup{\tagform@{\theequation}}}
\makeatother

\newcommand{\pa}{\partial}

\newtheorem*{definition*}{Definition}

\fontfamily{yfrak}

\begin{document}

\vskip 25mm

\begin{center}

{\large\bfseries



On the Fermionic Sector of Quantum Holonomy Theory

}

\vskip 6ex

Johannes \textsc{Aastrup}$^{a}$\footnote{email: \texttt{aastrup@math.uni-hannover.de}} \&
Jesper M\o ller \textsc{Grimstrup}$^{b}$\footnote{email: \texttt{jesper.grimstrup@gmail.com}}\\ 
\vskip 3ex

$^{a}\,$\textit{Mathematisches Institut, Universit\"at Hannover, \\ Welfengarten 1, 
D-30167 Hannover, Germany.}
\\[3ex]
$^{b}\,$\textit{QHT Gruppen, Copenhagen, Denmark.}
\\[3ex]

{\footnotesize\it This work is financially supported by Ilyas Khan, \\\vspace{-0cm}St. EdmundÕs College, Cambridge, United Kingdom and by\\ \vspace{-0,1cm}Tegnestuen Haukohl \& K\o ppen, Copenhagen, Denmark.}

\end{center}

\vskip 3ex

\begin{abstract}

In this paper we continue the development of quantum holonomy theory, which is a candidate for a fundamental theory based on gauge fields and non-commutative geometry. The theory is build around the $\mathbf{QHD}(M)$ algebra, which is generated by parallel transports along flows of vector fields and translation operators on an underlying configuration space of connections, and involves a semi-final spectral triple with an infinite-dimensional Bott-Dirac operator. Previously we have proven that the square of the Bott-Dirac operator gives the free Hamilton operator of a Yang-Mills theory coupled to a fermionic sector in a flat and local limit. In this paper we show that the Hilbert space representation, that forms the backbone in this construction, can be extended to include many-particle states.

\end{abstract}

\newpage

\section{Introduction}
\setcounter{footnote}{0}

In this paper we continue the development of {Quantum Holonomy Theory}, which is a candidate for a fundamental theory based on gauge fields and formulated within the framework of non-commutative geometry and spectral triples. 

The basic idea in Quantum Holonomy Theory is to start with an algebra that encodes the canonical commutation relations of a gauge theory in an integrated and non-local fashion. The algebra in question is called the quantum holonomy-diffeomorphisms algebra, denoted $\mathbf{QHD}(M)$, which was first presented in \cite{Aastrup:2014ppa} and which is generated by parallel transports along flows of vector fields and by translation operators on an underlying configuration space of gauge connections. In \cite{Aastrup:2015gba} it was demonstrated that this algebra encodes the canonical commutation relations of a gauge theory.

Once the $\mathbf{QHD}(M)$ has been identified the question arises whether it has non-trivial Hilbert space representations. This question was answered in the affirmative in \cite{Aastrup:2017vrm} where we proved that separable and strongly continuous Hilbert space representations of the $\mathbf{QHD}(M)$ exist in any dimensions. 
A key feature of these Hilbert space representations is that they are non-local. They are labelled by a scale $\tau$, which we tentatively interpret as the Planck scale and which essentially serves as a UV-regulator by suppressing modes in the ultra-violet. This UV-suppression does not break any spatial symmetries, i.e. these representations are isometric.
In \cite{Aastrup:2017atr} we constructed an infinite-dimensional Bott-Dirac operator that interacts with an algebra generated by holonomy-diffeomorphisms alone, denoted by $\mathbf{HD}(M)$, and proved that this Bott-Dirac operator together with the aforementioned Hilbert space representation forms a semi-finite spectral triple over a configuration space of connections.
In that paper we also demonstrated that the square of the Bott-Dirac operator coincides in a local and flat limit with the free Hamilton operator of a gauge field coupled to a fermionic sector, a result which opens the door to an interpretation of quantum holonomy theory in terms of a quantum field theory on a curved background.

In this paper we continue the analysis of these Hilbert space representations. One feature of the Bott-Dirac operator is that it naturally introduces the CAR algebra into the construction via an infinite-dimensional Clifford algebra. This CAR algebra has a natural interpretation in terms of a fermionic sector due to the aforementioned result that the square of the Bott-Dirac operator includes the Hamilton of a free fermion. One drawback of the Hilbert space representation constructed in \cite{Aastrup:2017vrm} is that it only involves what amounts to one-particle states. In other words, the Hilbert space representation does not act on the CAR algebra itself. In this paper we construct such a Hilbert space representation of the $\mathbf{QHD}(M)$ algebra. 
The result that such a representation exist solidifies the interpretation that quantum holonomy theory should be understood as a quantum theory of gauge fields coupled to fermions.  \\

This paper is organised as follows: We begin by introducing the $\mathbf{HD}(M)$ and $\mathbf{QHD}(M)$ algebras in section 2 and the infinite-dimensional Bott-Dirac operator in section 3. We then review the Hilbert space representation constructed in \cite{Aastrup:2017vrm}  in section 4. Finally we construct in section 5 a new Hilbert space representation where the $\mathbf{QHD}(M)$ algebra acts on the Fock space. We end with a discussion in section 6.

\section{The $\mathbf{HD}(M)$ and $\mathbf{QHD}(M)$ algebras}
\label{sektion2}

In this section we introduce the algebras $\mathbf{HD}(M)$ and $\mathbf{QHD}(M)$, which are generated by parallel transports along flows of vector-fields and for the latter part also by translation operators on an underlying configuration space of connections.  The $\mathbf{HD}(M)$ algebra was first defined in \cite{Aastrup:2012vq,AGnew} and the $\mathbf{QHD}(M)$ algebra in \cite{Aastrup:2014ppa}. In the following we shall define these algebras in a local and a global version. \\

Let $M$ be a compact manifold and let $\ca$ be a configuration space of gauge connections that takes values in the Lie-algebra of a compact gauge group $G$.  
A holonomy-diffeomorphism $e^X\in \mathbf{HD}(M)$ is a parallel transport along the flow $t\to \exp_t(X)$ of a vector field $X$.  
To see how this works we first let $\gamma$ be the path
$$\gamma (t)=\exp_{t} (X) (m) $$
running from $m$ to $m'=\exp_1 (X)(m)$. Given a connection $\nabla$ that takes values in a $n$-dimensional representation of the Lie-algebra $\mathfrak{g}$ of $G$  we then define a map
$$e^X_\nabla :L^2 (M )\otimes \mathbb{C}^n \to L^2 (M )\otimes \mathbb{C}^n$$
via the holonomy along the flow of $X$
\begin{equation}
  (e^X_\nabla \xi )(m')=    \hbox{Hol}(\gamma, \nabla) \xi (m)   ,
  \label{chopin1}
 \end{equation}
where $\xi\in L^2(M,\mathbb{C}^n)$ and where $\hbox{Hol}(\gamma, \nabla)$ denotes the holonomy of $\nabla$ along $\gamma$. This map gives rise to an operator valued function on the configuration space $\ca$ of $G$-connections via
\begin{equation}
\ca \ni \nabla \to e^X_\nabla  ,
\nn
\end{equation}
which we denote by $e^X$ and which we call a holonomy-diffeomorphism\footnote{The holonomy-diffeomorphisms, as presented here, are not a priori unitary, but by multiplying with a factor that counters the possible change in volume in (\ref{chopin1}) one can make them unitary, see \cite{AGnew}.}. For a function $f\in C^\infty (M)$ we get another operator valued function $fe^X$ on $\ca$.
We call the algebra generated by all holonomy-diffeomorphisms $e^X$ for the {\it global} holonomy-diffeomorphism algebra, denoted by $\mathbf{HD}_{\mbox{\tiny g}}(M)$, and we call the algebra generated by all holonomy-diffeomorphisms $f e^X$  for the {\it local} holonomy-diffeomorphism algebra, denoted simply by $\mathbf{HD}(M)$.\\

Furthermore, a $\mathfrak{g}$ valued one-form $\oo$ induces a transformation on $\ca$ and therefore an operator $U_\omega $ on functions on $\ca$ via   
$$
U_\omega (\xi )(\nabla) = \xi (\nabla - \omega) ,
$$ 
which gives us the quantum holonomy-diffeomorphism algebras, denoted either by $\mathbf{QHD}_{\mbox{\tiny g}}(M)$, which is the algebra generated by $\mathbf{HD}_{\mbox{\tiny g}}(M)$ and all the $U_\oo$ operators, or by $\mathbf{QHD}(M)$,  which is the algebra generated by $\mathbf{HD}(M)$ and all the $U_\oo$ operators (see also \cite{Aastrup:2014ppa}).

\section{An infinite-dimensional Bott-Dirac operator}
\label{Bott}

In this section we introduce an infinite-dimensional Bott-Dirac operator that acts in a Hilbert space that shall later play a key role in defining a representation of the $\mathbf{QHD}(M)$ algebras.
The following formulation of an infinite-dimensional Bott-Dirac operator is due to Higson and Kasparov \cite{Higson} (see also \cite{Aastrup:2017atr}).\\

Let $\ch_n= L^2(\mathbb{R}^n)$, where the measure is given by the flat metric, and consider the embedding
$$
\varphi_n : \ch_n\rightarrow\ch_{n+1}
$$
given by 
\begin{equation}
\varphi_n(\eta)(x_1,x_2,\ldots x_{n+1}) = \eta(x_1,\ldots, x_n)  \left(\frac{s_{n+1}}{\tau_2\pi}\right)^{1/4}e^{- \frac{s_{n+1} x_{n+1}^2}{2\tau_2}},
\label{ref}
\end{equation}
where $\{s_n\}_{n\in\mathbb{N}}$ is a monotonously increasing sequence of parameters, which we for now leave unspecified\footnote{In \cite{Higson} these parameters were not included, i.e. $s_n=1\forall n$.  }. This gives us an inductive system of Hilbert spaces 
$$
\ch_1\stackrel{\varphi_1}{\longrightarrow}  \ch_2 \stackrel{\varphi_2}{\longrightarrow}  \ldots   \stackrel{\varphi_n}{\longrightarrow}    \ch_{n+1} \stackrel{\varphi_{n+1}}{\longrightarrow}\ldots
$$
and we define\footnote{The notation $L^2 (\mathbb{R}^\infty )$, which we are using here, is somewhat ambiguous. We are here only considering functions on  $\mathbb{R}^\infty $ with a specific tail behaviour, namely the one generated by (3). We have not included this  tail behaviour in the notation. See \cite{Aastrup:2017vrm} for further details.}
 $L^2(\mathbb{R}^\infty) $ as the Hilbert space direct limit
\begin{equation}
L^2(\mathbb{R}^\infty) = \lim_{\rightarrow} L^2(\mathbb{R}^n)
\end{equation}
taken over the embeddings $\{\varphi_n\}_{n\in\mathbb{N}}$ given in (\ref{ref}).
%
We are now going to define the Bott-Dirac operator on $ L^2(\mathbb{R}^n)\otimes \Lambda^*\mathbb{R}^n$. 
Denote by $\mbox{ext}(v)$ the operator of external multiplication with $v$ on $\Lambda^*\mathbb{R}^n$, where $v$ is a vector in $\mathbb{R}^n$, and denote by $\mbox{int}(v)$ its adjoint, i.e. the interior multiplication by $v$.
Denote by $\{v_i\}$ a set of orthonormal basis vectors on $\mathbb{R}^n$ and let $\bar{c}_i$ and $c_i$ be the Clifford multiplication operators given by
\begin{eqnarray}
{c}_i &=& \mbox{ext}(v_i) + \mbox{int}(v_i)
\nn\\
\bar{c}_i &=& \mbox{ext}(v_i) - \mbox{int}(v_i) 
\end{eqnarray}
that satisfy the relations 
\begin{eqnarray}
 \{c_i, \bar{c}_j\} = 0, \quad
 \{c_i, {c_j}\} = \d_{ij}, \quad
 \{\bar{c}_i, \bar{c}_j\} =- \d_{ij}.
\end{eqnarray}
The Bott-Dirac operator on $ L^2(\mathbb{R}^n)\otimes \Lambda^*\mathbb{R}^n$ is given by
$$
B_n = \sum_{i=1}^n\left( \tau_2 \bar{c}_i   \frac{\pa}{\pa x_i} +  s_i c_i  x_i\right).
$$
With $B_n$ we can then construct the Bott-Dirac operator $B$ on $L^2(\mathbb{R}^\infty)\otimes \Lambda^*\mathbb{R}^\infty$ that coincides with $B_n$ on any finite subspace $L^2(\mathbb{R}^n)$. Here we mean by $\Lambda^*\mathbb{R}^\infty$ the inductive limit
$$
\Lambda^*\mathbb{R}^\infty= \lim_{\rightarrow} \Lambda^*\mathbb{R}^n.
$$
For details on the construction of $B$ we refer the reader to \cite{Higson} and to \cite{Aastrup:2017atr}, where we also showed that the square of $B$ coincides with the free Hamilton operator of a fermion Yang-Mills theory in a flat and local limit.

\section{A representation of the $\mathbf{QHD}(M)$ algebra}

In this section we write down the representation of the $\mathbf{QHD}(M)$ algebra, which was first constructed in \cite{Aastrup:2017vrm}. A key feature of this representation is that it involves a spatial non-locality characterised by a physical parameter $\tau_1$, which effectively acts as an ultra-violet regulator and which we in \cite{Aastrup:2017vrm} tentatively interpreted in terms of the Planck length.\\

To obtain a representation of the $\mathbf{QHD}(M)$ algebra we let 
$\langle \cdot\vert\cdot\rangle_{\mbox{\tiny s}} $ denote the Sobolev norm on $\OO^1(M\otimes\mathfrak{g})$, which has the form
\begin{equation}
\langle \omega_1\vert\omega_2\rangle_{\mbox{\tiny s}} 
:=
\int_M  \big( (1+ \tau_1\Delta^{\sigma})\omega_1  , (1+  \tau_1\Delta^{\sigma})\omega_2  \big)_{T_x^*M\otimes \mathbb{C}^n} (m) dm
\label{sob}
\end{equation}
where the Hodge-Laplace operator $\Delta$ and the  inner product  $(,)_{T_x^*M\otimes \mathbb{C}^n}$ on $T_x^*M\otimes \mathbb{C}^n$ depend on a metric g and where $\tau_1$ and $\sigma$ are positive constants. Also, we choose an $n$-dimensional representation of $\mathfrak{g}$.

Next, denote by $\{\xi_i\}_{i\in\mathbb{N}}$ an orthonormal basis of $\OO^1(M\otimes\mathfrak{g})$ with respect to the scalar product (\ref{sob}).
With this we can construct a space $L^2(\ca)$ as an inductive limit over intermediate spaces $L^2(\ca_n)$ with an inner product given by
\begin{eqnarray}
\langle \eta \vert \zeta \rangle_{\ca_n} &=& \int_{\mathbb{R}^n} \overline{\eta(x_1\xi_1 + \ldots + x_n \xi_n)} \zeta (x_1\xi_1 + \ldots + x_n \xi_n) dx_1\ldots dx_n \label{rn}
\end{eqnarray}
where $\eta$ and $\zeta$ are elements in $L^2(\ca_n)$, as explained in section 3, and also using the same tail behaviour as in section 3. Finally, we define the Hilbert space 
\begin{equation}
\ch_{\mbox{\bf\tiny YM}}= L^2(\ca)\otimes L^2(M, \mathbb{C}^n)
\label{ymm}
\end{equation}
in which we then construct the following representation of the $\mathbf{QHD}_{\mbox{\tiny l}}(M)$ algebra.

First, given a smooth one-form $\oo\in\OO^1(M,\mathfrak{g})$ we write $\oo =\sum \oo_i \xi_i$. The operator $U_\chi$  acts by translation in $L^2(\ca)$, i.e. 
\begin{eqnarray}
U_{\oo}(\eta) (\omega)&=&U_{\oo}(\eta) (x_1 \xi_1+x_2 \xi_2+ \ldots)
\nn\\
&=&  \eta ( (x_1+\oo_1)\xi_1+(x_2+\oo_2)\xi_2+ \ldots)  
\label{rep1}
\end{eqnarray}
with $\eta\in L^2(\ca)$. Next, we let $f e^X\in \mathbf{HD}(M)$ be a holonomy-diffeomorphism and $\Psi(\omega,m)=\eta(\omega)\otimes \psi(m)\in\ch_{\mbox{\tiny\bf YM}}$ where $\psi(m)\in L^2(M)\otimes \mathbb{C}^n$. We write
\begin{equation}
f e^X \Psi(\omega,m') =  f(m) \eta(\omega) Hol(\gamma, \omega) \psi(m)  
\label{rep2}
\end{equation}
where $\gamma$ is again the path generated by the vector field $X$ with $m'=\exp_1(X)(m)$.

In \cite{Aastrup:2017vrm} and \cite{Aastrup:2017atr} we prove that equations (\ref{rep1}) and (\ref{rep2}) gives a strongly continuous Hilbert space representation of the $\mathbf{QHD}(M)$ algebra in $\ch_{\mbox{\bf\tiny YM}}$.
Note that this representation is isometric with respect to the background metric $g$, see \cite{Aastrup:2017vrm} for details.

\section{Representing $\mathbf{QHD}_{\mbox{\tiny g}}(M)$ on the Fock Space}

The Bott-Dirac operator acts on $L^2 (\mathbb{R}^\infty )\otimes \Lambda^*\mathbb{R}^\infty$, and not on $L^2(\ca)\otimes L^2(M,\mathbb{C}^n)$ as the  $\mathbf{QHD}(M)$-algebra does. The Hilbert space $L^2(\mathbb{R}^\infty)$ is, however, easily identified with $L^2 (\ca )$ via  
$$\mathbb{R}^n \ni (x_1,\ldots , x_n) \mapsto x_1\xi_1 +\ldots + x_n \xi_n \in \ca_n . $$
We will therefore denote  $\Lambda^*\mathbb{R}^\infty$ by  $\Lambda^*\ca$. We thus get an action of the Bott-Dirac operator and the $\mathbf{QHD}(M)$-algebra on  $L^2(\ca)\otimes\Lambda^*\ca \otimes L^2(M,\mathbb{C}^n)$. This is  somewhat unsatisfactory due to two reasons:
\begin{enumerate}
\item The Fermions on which the $\mathbf{QHD}(M)$-algebra acts, is a one-particle space. We could of course try to take the Fock space of $L^2(M,\mathbb{C}^n)$ instead of just $L^2(M,\mathbb{C}^n)$.
\item We have a fermionic doubling in the sense that we have the fermionic Fock space $\Lambda^*\ca$, where the bosons, i.e. the $\mathbf{QHD}(M)$-algebra, do not act at all, and then the fermions in $L^2(M,\mathbb{C}^n)$, where the bosons do act.
\end{enumerate} 
It is therefore desirable to get an action the $\mathbf{QHD}(M)$ algebra on $L^2 (\ca )\otimes \Lambda^* \ca $. In this section we show how this can be accomplished for the $\mathbf{QHD}_{\mbox{\tiny g}}(M)$ algebra but at the present moment not for the local $\mathbf{QHD}(M)$ algebra.\\

We begin with the basespace 
\begin{equation}
\mathcal{H}^\sigma =\Omega^1 (M,\mathfrak{g}) ,
\label{trmpp}
\end{equation}
where the Hilbert space structure is again with respect to a suitable Sobolev norm (\ref{sob})
in the sense that the righthand side of (\ref{trmpp}) has been completed in this norm (we remind the reader that the superscript '$\sigma$' is the power of the Laplace operator in (\ref{sob})). 
The main purpose is to get a unitary, connection dependent action of the group of holonomy-diffeomorhphisms in $\mathbf{HD}_{\mbox{\tiny g}}(M)$ on the Hilbert space $\mathcal{H}^\sigma$. Once we have a unitary action it extends uniquely to an action on the associated Fock space  $\Lambda^* \mathcal{H}^\sigma$  via
 $$F_\nabla(v_1\wedge \ldots \wedge v_n)=F_\nabla (v_1)\wedge \ldots \wedge F_\nabla (v_n) , $$
 where $F$ denotes a holonomy-diffeomorphism and $\nabla$ denotes a connection.  
Once we have this we get a unitary action of the $\mathbf{HD}_{\mbox{\tiny g}}(M)$ algebra on $\Lambda^* \mathcal{H}^\sigma \otimes L^2(\mathcal{A})$ via 
$$ F (\xi \otimes \eta )(\nabla ) =F_\nabla  (\xi)\eta ( \nabla ) .$$

The question is of course how we get an action of  
 $\mathbf{HD}_{\mbox{\tiny g}}(M)$ on $\mathcal{H}^\sigma$. To answer this question we let $F$ be a holonomy-diffeomorphism and let $\nabla$ be a  $\mathfrak{g}$-connection. We start with the case $\sigma=0$. Let $F$ be the flow of the vector field $X$ and let $\omega \in  \Omega^1 (M,\mathfrak{g})$. 
Let $m_1\in M$ and $m_2=\exp (X)(m_1)$, and $\gamma$ the path $t\to e^{tX}(m_1)$. Furthermore we denote by $(e^{-X})^* (\omega )$ the pullback of the one-form part of $\omega$ by the diffeomorphism $e^{ -X}$, i.e. $(e^{-X})^*$ leaves the Lie algebra $\mathfrak{g}$ unchanged.  We define
$$e_\nabla^{X}(\omega ) (m_2)= \hbox{Hol}  (\gamma ,\nabla)\Big( (e^{-X})^*(\omega)(m_2) \Big)  (\hbox{Hol} (\gamma ,\nabla))^{-1} .$$

This does not define a unitary operator, unless $\exp (X)$ is an isometric flow. Unlike in section \ref{sektion2}  we cannot adjust the lack of unitarity by multiplying by a suitable determinant. The problem lies in the one form part. One possible way to deal with this is to consider only holonomy-diffeomorphisms, which are isometries with respect to a chosen metric. Alternatively -- and this is the option that we shall adopt -- we can allow the operators to be non-unitary. In this latter case we will still get bounded operators on $\mathcal{H}^\sigma$, even when we consider the supremum over all connections. The problem is,  that when we extend the action to the Fock space the operators will no longer be bounded. The unboundedness is however not so severe since the operators are bounded when we consider them only acting on a subspace of the Fock space which contains particle states with particle number bounded by a given value.

For general $\sigma$'s there is a natural way to proceed: The map $1+\tau_1\Delta^\sigma: \mathcal{H}^0 \to \mathcal{H}^\sigma$ is a unitary operator, and to get the action on $\mathcal{H}^\sigma$ we simply conjugate the action we have on $\mathcal{H}^0 $ with $1+\tau_1\Delta^\sigma$. If we choose holonomy-diffeomorphisms, which are isometries, this gives a unitary action. For general $\sigma$'s, we could also just proceed directly like above, without conjugating with $1+\tau_1\Delta^\sigma$.   However without this conjugation the action would not be unitary on $\mathcal{H}^\sigma$, $\sigma\not= 0$, for the isometric flows. \\

Finally, this representation of the $\mathbf{HD}_{\mbox{\tiny g}}(M)$ algebra is straight forwardly extended to the full $\mathbf{QHD}_{\mbox{\tiny g}}(M)$ algebra via
\begin{equation}
 U_\oo (\xi \otimes \eta )(\nabla ) = (\xi \otimes \eta )(\nabla +\oo)  ,
 \label{NOO2}
 \end{equation}
for $\xi\otimes\eta\in \Lambda^* \mathcal{H}^\sigma \otimes L^2(\mathcal{A})$. 
This implies that we have a non-unitary action of the $\mathbf{QHD}_{\mbox{\tiny g}}(M)$ algebra on $\Lambda^*\mathcal{A}\otimes L^2 (\mathcal{A})$. The action is strongly continuous if it is restricted to finite particle states.

Note that the reason that this representation does not work for the local $\mathbf{HD}(M)$ and $\mathbf{QHD}(M)$ algebras is that it is not clear what the action of a function $f\in C^\infty(M)$ should be on the $0$-forms in $\Lambda^* \mathcal{H}^\sigma$, i.e. on the vacuum. For this reason we leave out the $C^\infty(M)$ part and consider instead only global holonomy-diffeomorphisms.

\section{Discussion}

In this paper we show that the representation of the $\mathbf{QHD}(M)$ algebra constructed in \cite{Aastrup:2017vrm} can be extended to include also the CAR algebra if we consider only global holonomy-diffeomorphisms. 
This result provides us with what we believe is a completely new interpretation of  fermionic quantum field theory in terms of geometrical data of  a configuration space of connections. Consider first the ordinary Dirac operator and a spin-geometry. Here the fermion can be viewed as being part of an encoding of geometrical data of the underlying manifold, i.e. a spectral triple. In our case we have instead of the 4-dimensional Dirac operator an infinite dimensional Bott-Dirac operator acting in a Hilbert space over a configuration space of connections. This means that the CAR algebra and the fermionic sector is part of an encoding of geometrical data of this configuration space.  \\

As we demonstrated in \cite{Aastrup:2017atr} quantum holonomy theory is closely related to quantum field theory, the latter being  based on two basic principles: locality and Lorentz invariance. In the axiomatic approaches these principles are encoded in the Osterwalder Schrader \cite{Osterwalder:1973dx} axioms for the Euclidean theory and in the G\.{a}rding-Wightman \cite{Wightman} or the Haag-Kastler \cite{Haag:1963dh} axioms for the Lorentzian theory. 
To understand the difference between the present approach and ordinary quantum field theory we need to understand the role of the ultra-violet regulator in the form of the Sobolev norm (\ref{sob}), which is the central element required to secure the existence of the Hilbert space representations. There are two options: either this regulator is a traditional cut-off that should eventually be taken to zero or it is a physical feature of this particular theory.

If the ultra-violet regulator is a traditional cut-off then we are firmly within the boundaries of ordinary quantum field theory albeit with a different approach and with a different toolbox. In that case the question is whether the introduction of the Bott-Dirac operator and the fact that we have a spectral triple will give us new information about the limit where the cut-off goes to zero. Similar to algebraic quantum field theory \cite{Haag:1992hx} this approach is not limited in its choice of background.

If on the other hand the regulator is to be viewed as a physical feature then we are decidedly outside the realm of traditional quantum field theory. There are two immediate consequences:
\begin{enumerate}
\item The Lorenz symmetry will be broken. The Hilbert space representation based on the Sobolev norm (\ref{sob}) is isometric with respect to the metric on the three-dimensional manifold but the Lorentz symmetry will not be preserved. Instead there will be a larger symmetry that involves a scale transformation. 
\item
The theory is non-local. Whereas ordinary quantum field theory is based on operator valued distributions the present setup does not permit sharply localised entities. This also implies that the canonical commutation relations will only be realised up to a correction at the scale of the regularisation. 
\end{enumerate}
Clearly this breaks with all the aforementioned axiomatic systems but the question is whether it is physically feasible? We believe that it is. First of all, it is not known whether the Lorentz symmetry is an exact symmetry in Nature and indeed much experimental effort has gone into testing whether it is \cite{Jacobson:2004rj}. We believe that the experimental constraints are sufficiently weak to permit the type of Lorenz breaking that we propose as long as it is restricted to the Planck scale. Secondly, it is generally believed that exact locality is not realised in Nature. Simple arguments combining quantum mechanics with general relativity strongly suggest that distances shorter than the Planck length are operational meaningless \cite{Doplicher:1994tu}. It is generally believed that a Planck scale screening will be produced by a theory of quantum gravity but we see no reason why it cannot be generated by quantum field theory itself as a part of its representation theory. We would then of course need to address the question of which regulator to choose, since the regulator would now be a quantity, that in principle is obvervable.

\vspace{1cm}
\noindent{\bf\large Acknowledgements}\\

\noindent
JMG would like to express his gratitude towards Ilyas Khan, United Kingdom, and towards the engineering company Tegnestuen Haukohl \& K\o ppen, Denmark, for their generous financial support. JMG would also like to express his gratitude towards the following sponsors:  Ria Blanken, Niels Peter Dahl, Simon Kitson, Rita and Hans-J\o rgen Mogensen, Tero Pulkkinen and Christopher Skak for their financial support, as well as all the backers of the 2016 Indiegogo crowdfunding campaign, that has enabled this work. Finally, JMG would like to thank the mathematical Institute at the Leibniz University in Hannover for kind hospitality during numerous visits.\\


\begin{thebibliography}{99}






\bibitem{Aastrup:2014ppa}
  J.~Aastrup and J.~M.~Grimstrup,
  ``The quantum holonomy-diffeomorphism algebra and quantum gravity,''
  Int.\ J.\ Mod.\ Phys.\ A {\bf 31} (2016) no.10,  1650048.





\bibitem{Aastrup:2015gba}
  J.~Aastrup and J.~M.~Grimstrup,
  ``Quantum Holonomy Theory,''
  Fortsch.\ Phys.\  {\bf 64} (2016) no.10,  783.





\bibitem{Aastrup:2017vrm}
  J.~Aastrup and J.~M.~Grimstrup,
  ``Representations of the Quantum Holonomy-Diffeomorphism Algebra,''
  arXiv:1709.02943.





\bibitem{Aastrup:2017atr}
  J.~Aastrup and J.~M.~Grimstrup,
  ``Nonperturbative Quantum Field Theory and Noncommutative Geometry,''
  arXiv:1712.05930.

  
  
  
  

\bibitem{Aastrup:2012vq}
  J.~Aastrup and J.~M.~Grimstrup,
  ``C*-algebras of Holonomy-Diffeomorphisms and Quantum Gravity I,''
  Class.\ Quant.\ Grav.\  {\bf 30} (2013) 085016.



  
  

\bibitem{AGnew}
  J.~Aastrup and J.~M.~Grimstrup,
  ``C*-algebras of Holonomy-Diffeomorphisms and Quantum Gravity II'',
   J.\ Geom.\ Phys.\  {\bf 99} (2016) 10.

















  
\bibitem{Higson}
N.~Higson and G.~Kasparov, "E-theory and KK-theory for groups which act properly and isometrically on Hilbert space", 
Inventiones Mathematicae, vol. {\bf 144}, issue 1, pp. 23-74.






 














  
  


  

 







\bibitem{Osterwalder:1973dx}
  K.~Osterwalder and R.~Schrader,
  ``Axioms For Euclidean Green's Functions,''
  Commun.\ Math.\ Phys.\  {\bf 31} (1973) 83.


\bibitem{Wightman}
A.~S.~Wightman,  ``HilbertÕs sixth problem: Mathematical treatment of the axioms of physics ``, in F.E. Browder (ed.): Mathematical Developments Arising from HilbertÕs Problems, Vol. 28:1 of Proc. Symp. Pure Math., Amer. Math. Soc, 1976, pp. 241 - 268.



\bibitem{Haag:1963dh}
  R.~Haag and D.~Kastler,
  ``An Algebraic approach to quantum field theory,''
  J.\ Math.\ Phys.\  {\bf 5} (1964) 848.




\bibitem{Haag:1992hx}
  R.~Haag,
  ``Local quantum physics: Fields, particles, algebras,''
  Berlin, Germany: Springer (1992) 356 p. (Texts and monographs in physics).

\bibitem{Jacobson:2004rj}
  T.~Jacobson, S.~Liberati and D.~Mattingly,
  ``Astrophysical bounds on Planck suppressed Lorentz violation,''
  Lect.\ Notes Phys.\  {\bf 669} (2005) 101.




\bibitem{Doplicher:1994tu}
  S.~Doplicher, K.~Fredenhagen and J.~E.~Roberts,
  ``The Quantum structure of space-time at the Planck scale and quantum fields,''
  Commun.\ Math.\ Phys.\  {\bf 172} (1995) 187.






\end{thebibliography}
\end{document}